\documentclass[11pt,twoside]{article}

\usepackage{asp2006}
\usepackage{epsf}
\usepackage{lscape}
\usepackage{times}

\begin{document}
\title{Collateral Damage: the Implications of Utrecht Star Cluster Astrophysics for Galaxy Evolution}
\author{J.~M.~Diederik~Kruijssen}
\affil{Max-Planck Institut f\"{u}r Astrophysik, Karl-Schwarzschild-Stra{\ss}e 1, 85748, Garching, Germany; {\tt kruijssen@mpa-garching.mpg.de}}

\begin{abstract}
Until the early 2000s, the research portfolio of the Astronomical Institute in Utrecht (SIU) did not include galaxy evolution. Somewhat serendipitously, this changed with the advent of the star cluster group. In only a few years, a simple framework was developed to describe and quantify the properties of dynamically evolving star cluster populations. Since then, the `Utrecht cluster disruption model' has shown that the galactic environment plays an important role in setting the evolution of stellar clusters. From this simple result, it follows that cluster populations bear some imprint of the characteristics and histories of their host galaxies, and that star clusters can be used to trace galaxy evolution -- an aim for which the Utrecht star cluster models were never designed, but which they are well-capable of fulfilling. I review some of the work in this direction, with a strong emphasis on the contributions from the SIU.
\end{abstract}

In 1920, the famous Dutch poet Adriaan Roland Holst (1888--1976) published the wonderful story {\it Deirdre en de zonen van Usnach},\footnote{{\it Deirdre and the Sons of Usnach}.} based on an Irish-Celtic myth from the Ulster Cycle. It tells the tale of the lady Deirdre, who is so incredibly beautiful that every man who sees her loses his heart. While this enables her to do a lot of good, the story does not end well. In the finale, Deirdre goes west, to the seashore, where far beyond the horizon the paradise of Elysium lies. As she wanders into the sea and lets the waves take her, she leaves this world -- but not without having changed it forever.

The closure of the Astronomical Institute Utrecht (SIU) prompted the conference of these proceedings, held at the Dutch west coast and aimed at celebrating the achievements of 370 years of astronomy in Utrecht. The occasion and location of the meeting drew a striking parallel with Deirdre's fate. Fortunately none of the participants followed her example and went for a swim -- the North Sea is quite cold in early April.

\section{Introduction}
A stellar cluster evaporates due to successive encounters between its stars. The rate at which this process proceeds is not only set by internal dynamics, but also by the tidal field the cluster resides in. The stronger the tidal field is, the smaller the Jacobi radius beyond which stars are unbound, and hence the more easily stars can escape \citep[e.g.][]{baumgardt01}. If the orbital motion of a cluster causes it to experience time-dependent tidal forces, a second environmental mechanism for cluster disruption comes into play. Varying tidal fields induce tidal heating, which increases the kinetic energy of the stars \citep{ostriker72}. Examples in nature of these `tidal shocks' are pericentre passages or galaxy disc crossings of globular clusters \citep[e.g.][]{gnedin97}, encounters with giant molecular clouds \citep{gieles06}, and spiral arm passages \citep{gieles07}. These examples illustrate that the disruption rate of stellar clusters is largely set by the galactic environment. In practice, the described cluster destruction processes are effectively mass-dependent, in that low-mass clusters are more rapidly disrupted than massive ones.

Due to the relation between the galactic environment and the dynamical evolution of stellar clusters, the properties of cluster populations vary among different galaxies. High-density galaxies with correspondingly high disruption rates \citep{lamers05a,kruijssen11} generally contain very few old clusters \citep{boutloukos03,bastian05,gieles05}. Additionally, the mass-dependence of cluster disruption implies that disruptive environments contain fewer low-mass clusters \citep{larsen09,gieles09}. These trends are also present in a spatially resolved sense within single galaxies \citep{gieles05,kruijssen11,bastian11,bastian12}, as well as over the course of galaxy histories, during which the galactic environment may have experienced substantial evolution \citep{kruijssen12c}. When interpreted correctly, these relations imply that cluster populations provide clues to the evolutionary histories of their host galaxies.

\section{The galactic environment and the Utrecht cluster disruption model}
\citet{lamers05} developed a simple analytic model to describe the age and mass distributions of evolving star cluster populations. In this model, the departure of both distributions from their initial forms (the cluster formation history and the initial cluster mass function, respectively) is caused by a non-zero disruption rate (or a finite disruption timescale) that is set by the galactic environment. For each galaxy individually, this rate has been assumed to be constant in space and time. The power of this single-parameter approach has been demonstrated by various successful applications to the cluster populations of nearby galaxies \citep{bastian05,gieles05,lamers06a,gieles08b}.

It was shown by \citet{boutloukos03} that the disruption rates of clusters vary between the Small Magellanic Cloud (SMC), the solar neighbourhood, M33, and M51. \citet{lamers05a} then found a correlation between the disruption rate and the ambient density within the host galaxy, approximately following the theoretically predicted relations by \citet{portegieszwart98} and \citet{baumgardt03}. The one exception was the interacting galaxy M51, which turned out to have a disruption rate that is an order of magnitude higher than expected. We later showed in \citet{kruijssen12c} that such an increase can be explained by considering the growth of the gas density during galaxy interactions, and the correspondingly enhanced disruption rate due to tidal shocks. Similarly, the single-parameter approach of a disruption rate that is constant in time and space does not successfully describe the cluster population of the interacting Antennae galaxies \citep{gieles08b}.

While the above examples clearly show that the galactic environment influences cluster disruption and hence is an essential factor in setting the properties of the cluster population, it is also evident that the assumption of a single disruption rate throughout space and time only holds for a certain subset of (quiescent) galaxies. This assumption has recently been overcome in \citet{kruijssen11}, where we expanded the Utrecht disruption model by including it in numerical simulations of galaxy evolution, in which the disruption rate is determined for each cluster individually, accounting for the steady tidal field as well as for tidal shocks. The numerical models self-consistently follow the impact of the galactic environment on cluster disruption, and have been used to show that encounters with giant molecular clouds are the dominant disruption mechanism in disk galaxies \citep[also see][]{gieles06,lamers06a}. They have also been applied to predict how the cluster age distribution changes under different galactic conditions, providing some characteristic features that can be verified in observational work. For instance, we have shown that the mean disruption rate of a cluster population decreases with age \citep[also see][]{elmegreen10b}, due to the preferential survival of clusters in quiescent environments (`natural selection') and their migration away from their dense, natal environment. Observational evidence for such variations has been found by \citet{bastian11,bastian12}.

\section{Star cluster populations as tracers of galaxy evolution}
There are numerous ways in which the properties of the star cluster population may be used to derive the galaxy evolutionary history. In this section, we discuss some examples and provide an outlook for how this avenue can be expanded in future work.

The decrease of the disruption rate with age affects the slope of the cluster age distribution over the age range in which cluster migration and natural selection act on the cluster population. Both effects are most important in galaxies with a high density contrast between star-forming regions and their surroundings \citep{kruijssen11}. Because migration occurs on the global dynamical timescale of a system, it can be used to infer the duration of certain events in the history of the galaxy. For instance, it indicates the interaction timescale of an ongoing galaxy collision, or the dynamical timescale of an isolated galaxy. The latter tracer can be verified using an example. The dynamical timescale of the disc galaxy M83 is $1/\Omega\sim20$~Myr, and there are indications of an evolving disruption rate over the same age range \citep{bastian11,bastian12}.

Another illustration of how the galactic environment affects the cluster population was found in \citet{kruijssen12c}, where we performed numerical simulations of major mergers of disc galaxies, again including a model for the evolution of the cluster population. In galaxy mergers, tidal torques drive the gas towards the galaxy centres. As a result, the disruption of clusters by dense giant molecular clouds \citep{gieles06} is prevalent, and counteracts a simultaneous increase of the cluster formation rate. The number of clusters in the merger remnant is therefore lower than in the progenitor disc galaxies. This can be quantified as a `survival cluster fraction', which we have found to decrease with the starburst intensity (or increase with the gas depletion timescale). Number counts of stellar clusters in merger remnants could thus be used to infer the characteristics of the preceding galaxy interaction and starburst.

Extrapolating the approach of tracing galaxy-scale events using the surviving cluster population, old globular clusters are exemplary targets for similar analyses, because they may be used to infer the conditions of galaxy formation at high redshift. Although they have substantially higher characteristic masses than young clusters in nearby galaxies, it has been argued that globular clusters formed through the same physical mechanisms that we see locally \citep{elmegreen97,kruijssen12b}. Their high average mass then implies that the vast majority of lower-mass globular clusters must have been disrupted due to dynamical evolution. However, the degree of disruption appears to be universal, which is hard to understand considering the widely varying galactic environments in which globular clusters currently reside. The efficient destruction of clusters in dense, star-forming environments might provide a solution. Globular clusters all formed in the dense, high-redshift universe, and if most of their disruption occurred early on, their present-day properties should be similar \citep{elmegreen10,kruijssen12c}. If true, this puts constraints on the high-redshift conditions under which globular clusters formed.

The origin of globular clusters is far from being a solved problem. The above scenario is a plausible solution, but it is generalised from physical systems that may not be adequate. It is thought that globular clusters did not form in major mergers, but instead have their roots in starburst dwarf galaxies, unstable high-redshift discs, and other high-redshift, star-forming environments (see \citealt{kruijssen12c} and references therein). While it seems reasonable that these conditions were equally disruptive as the starbursts in our galaxy merger simulations, this requires explicit verification. Numerical simulations of galaxy formation with cosmologically motivated initial conditions \citep[e.g.][]{prieto08} will be necessary to provide a definitive answer.

\acknowledgements
I am very grateful to the former SIU staff -- Henny Lamers in particular -- for their unlimited dedication to science and teaching.

\end{document}